%% file: GCpaper.tex
\documentclass[conference]{IEEEtran}
\IEEEoverridecommandlockouts
\usepackage{cite}
\usepackage[T1]{fontenc}
\usepackage{amsmath,amssymb,amsfonts,amsthm}
\usepackage{algorithm}
\usepackage{algpseudocode,algorithmicx}
\usepackage{graphicx}
\graphicspath{{./figures/}}
\usepackage{textcomp}
\usepackage{xcolor}
\usepackage{balance}
\usepackage[unicode=true,
 bookmarks=false,bookmarksnumbered=true,bookmarksopen=false,bookmarksopenlevel=1,
 breaklinks=false,pdfborder={0 0 0},pdfborderstyle={},backref=false,colorlinks=false]
 {hyperref}
\hypersetup{pdftitle={Title},
 pdfauthor={Admin},
 pdfpagelayout=OneColumn, pdfnewwindow=true, pdfstartview=XYZ, plainpages=false}

\makeatletter

\DeclareMathOperator*{\argmin}{arg\,min\ }
\DeclareMathOperator*{\argmax}{arg\,max\ }
\theoremstyle{remark}

\let\saved@hyper@linkurl\hyper@linkurl
\let\saved@hyper@link@\hyper@link@
\AtBeginDocument{%
  \NoHyper
  \let\hyper@linkurl\saved@hyper@linkurl 
  \let\hyper@link@\saved@hyper@link@ 
}
\makeatother

\begin{document}

\title{\huge Gridless Angular Domain  Channel Estimation for mmWave Massive MIMO System With One-Bit Quantization Via Approximate Message Passing 
}

\author{\IEEEauthorblockN{ Liangyuan~Xu\textsuperscript{$\S \ast$}, Feifei~Gao\textsuperscript{$\S \ast$}, Cheng~Qian\textsuperscript{$\dag $}}
\IEEEauthorblockA{{\textsuperscript{$\S $} Department of Automation, Tsinghua University, Beijing, China} \\
{\textsuperscript{$\ast $} Beijing National Research Center for Information Science and Technology (BNRist)}\\
{\textsuperscript{$\dag $} Department of Electrical and Computer Engineering, University of Virginia, Charlottesville, USA } \\
e-mail: \protect\href{mailto:xly18@mails.tsinghua.edu.cn}{xly18@mails.tsinghua.edu.cn}; \protect\href{mailto:feifeigao@ieee.org}{feifeigao@ieee.org}; \protect\href{mailto:alextoqc@gmail.com}{alextoqc@gmail.com} 
}
}
\maketitle

\begin{abstract}
We develop a direction of arrival (DoA) and channel estimation algorithm for the one-bit quantized millimeter-wave (mmWave) massive multiple-input multiple-output (MIMO) system. By formulating the estimation problem as a noisy one-bit compressed sensing problem, we propose a computationally efficient gridless solution based on the  expectation-maximization generalized approximate message passing (EM-GAMP) approach. The proposed algorithm does not need the prior knowledge about the number of DoAs  and  outperforms the existing methods in distinguishing extremely close DoAs for the case of one-bit  quantization. Both the DoAs and the channel coefficients are estimated for the case of one-bit  quantization. The simulation results show that the proposed  algorithm has effective estimation performances when the DoAs are very close to each other.
\end{abstract}


\section{Introduction}

Massive multi-input multi-output (MIMO), a promising  technology for 5G mobile communication systems, deploys hundreds of antennas at base stations (BSs) and has intrinsic capabilities to estimate as well as leverage direction of arrival (DoA) \cite{MIMODoA}. On the other side, in the millimeter-wave (mmWave) frequency band, it is possible to pack a large number of antennas in a small area, which makes massive MIMO attractive. However, the hardware cost and power consumption will increase significantly as the dimensionality of massive MIMO systems grows, especially the power consumption of analog-to-digital converters (ADCs). This issue motivates the studies  for 1-bit quantized mmWave massive MIMO systems.

Compressed sensing (CS) based methods for DoA estimation with 1-bit quantization have been investigated in \cite{BIHTDoA,SVMDoA,WenDoA}. In \cite{BIHTDoA}, the binary iterative hard thresholding (BIHT) algorithm from \cite{BIHT} was extended and applied to  1-bit DoA estimation. However, the approach in \cite{BIHTDoA} is on-grid based and would suffer from the grid-mismatch problem. The authors of \cite{SVMDoA} proposed an off-grid solution by using Taylor-expansion to refine the grid. The work \cite{WenDoA} provided a gridless estimation method based on the atomic norm and the Vandermonde decomposition.  In \cite{1bRELAX}, 1-bit RELAX algorithm which extended the maximum likelihood (ML) estimator was utilized.  However, this method needs iterative 2-dimensional coarse searches, which is computationally prohibitive and impractical. In addition, the above studies need to know the number of DoAs prior to the estimation.

When the DoAs are very close, the DoA estimation would be a challenging task for both 1-bit  and $\infty$-bits quantization. All the 1-bit DoA estimation methods proposed in \cite{BIHTDoA,SVMDoA,1bRELAX}  have problems distinguishing extremely close DoAs. Even for the ideal case of $\infty$-bits quantization, the DoA estimation algorithm in \cite{WBL} and  the spatial basis expansion method in \cite{xhxTVT} would suffer severe performance degradation in distinguishing very close DoAs. Hence, this issue need to be addressed. 

In this paper, we investigate DoA estimation and channel estimation of massive MIMO system with 1-bit quantization. By exploiting the sparsity in the angular domain and formulating the estimation problem as  a noisy 1-bit compressed sensing problem, we design a computationally efficient gridless estimation method based on the expectation-maximization generalized approximate message passing (EM-GAMP) approach \cite{EMGAMP}.  By decoupling all the DoAs and estimating them separately, our method surpasses the existing algorithms when distinguishing extremely close DoAs.

\section{System Model}

\begin{figure}[!tp]
\centering
\includegraphics[width=0.4\textwidth]{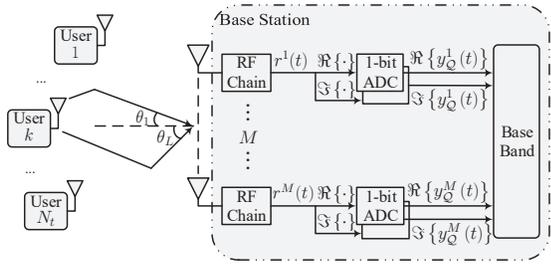}
\caption{Block diagram of 1-bit quantized multi-user massive MIMO system}
\label{fig:systdiagram}
\end{figure}

Consider an uplink mmWave massive MIMO system where $N_t$ single antenna users are served by a BS with uniform linear array (ULA) of $M$ antennas and 1-bit quantization, as demonstrated in Fig. \ref{fig:systdiagram}. Let the inter-element space of the ULA be half the wavelength. At time instant $t$, the received signal vector ${\bf r} \left( t \right) \in \mathbb{C}^{M} \times 1$ at the BS is given by 
\begin{equation}\label{eq:yvec}
\mathbf{r}(t)=\sum\limits_{k=1}^{N_t}{\sum\limits_{l=1}^{L_{k}}{\alpha_{l, k} \mathbf{a}_{r}\left(\theta_{l, k}\right) s_{k}(t)} }  + \mathbf{n},
\end{equation}
where $L_{k}$ is the number of paths between the $k$th user and the BS, $\alpha_{l, k}$ and $\theta_{l, k}$ represent the channel complex gain and the DoA for the $l$th path of the $k$th user respectively, $\mathbf{a}_{r}\left(\theta_{l, k}\right) = \left[ {1,{e^{ - \jmath \pi \sin \left( {\theta _{l,k}} \right)}}, \cdots ,{e^{ - \jmath \left( M - 1 \right)\pi \sin \left( {\theta _{l,k}} \right)}}} \right]^T$ denotes the steering vector in direction ${\theta _{l,k}}$ at the BS, $\jmath = \sqrt{-1}$, $s_{k}(t)$ is the symbol transmitted by the $k$th user,  and $\mathbf{n} \sim  \mathcal{C N}\left(\mathbf{n}; \mathbf{0}, \sigma^2\mathbf{I}\right)$ is the additive white Gaussian noise vector. 

Rewriting \eqref{eq:yvec} into matrix form yields
\begin{equation}\label{eq:ymtx}
{\mathbf{r}}\left( t \right) = {\bf{V}}\left( {\boldsymbol{\theta }} \right){\bf{Ds}}\left( t \right) + {\bf{n}},
\end{equation}
where $ {\bf{V}}\left( {\boldsymbol{\theta }} \right) $ is the $ M \times \sum\nolimits_{k = 1}^{{N_t}} {L_k} $ steering matrix with Vandermonde structure ${\bf{V}}\left( {\boldsymbol{\theta }} \right) = \left[ {{{\bf{a}}_r}\left( {{\theta _{1,1}}} \right), \cdots ,{{\bf{a}}_r}\left( {{\theta _{L_{N_t},N_t}}} \right)} \right]$, ${\boldsymbol{\theta }} = [\theta_{1,1},\cdots,\theta_{L_{N_t},N_t}]$, ${\bf{D}} = {\rm{diag}}\left\{ {{\alpha _{1,1}}, \cdots ,{\alpha _{L_{N_t},N_t}}} \right\}$ is a $\sum\nolimits_{k = 1}^{{N_t}} {L_k} \times \sum\nolimits_{k = 1}^{{N_t}} {L_k}$ diagonal matrix, and ${\bf{s}}\left( t \right)$ denotes the $\sum\nolimits_{k = 1}^{{N_t}} {L_k} \times 1$ symbol vector. We define ${\mathbf H} \triangleq {\bf{V}}\left( {\boldsymbol{\theta }} \right){\bf{D}}$ as the effective channel.

The discrete Fourier transform (DFT) of the effective channel is
\begin{equation}\label{eq:effCH}
{\bf X}\left( {\boldsymbol{\theta }} \right) \triangleq {\bf{F}}{\mathbf H}={\bf{F}}{\bf{V}}\left( {\boldsymbol{\theta }} \right){\bf{D}},
\end{equation}
where ${\bf{F}}$ is the $M\times M$ normalized DFT matrix with ${{{\bf{F}}^H}\bf{F}}={\mathbf I}_{M}$. For mmWave massive MIMO, it has been previously discussed in \cite{xhxTVT,WBL} that  each column of ${\bf X}\left( {\boldsymbol{\theta }} \right)$ is sparse due to the special Vandermonde structure of ${\bf{V}}\left( {\boldsymbol{\theta }} \right)$, e.g., a large portion of the elements of each column are very close to zero. The sparsity of ${\bf X}\left( {\boldsymbol{\theta }} \right)$ will be exploited in the proposed channel estimation method. 

After 1-bit quantization, the received signal should be rewritten as
\begin{equation}
\mathbf{y}_{\mathcal{Q}}(t)={\mathcal Q}\left( {{\bf{r}}\left( t \right)} \right) = {\operatorname{ sign}} \left( {\Re \left\{ {{\bf{r}}\left( t \right)} \right\}} \right) + \jmath {\operatorname{ sign}} \left( {\Im \left\{ {{\bf{r}}\left( t \right)} \right\}} \right),
\end{equation}
where ${\mathcal Q}\left( \cdot \right)$ is the 1-bit quantizer, ${\mathop{\rm sign}\nolimits} \left( \cdot \right)$ is the signum function, and both ${\mathcal Q}\left( \cdot \right)$ and ${\mathop{\rm sign}\nolimits} \left( \cdot \right)$ are applied component-wise and separately to the real and imaginary parts of matrices. Hence, the received 1-bit quantized baseband signals at BS can be expressed as 
\begin{equation}\label{eq:qyvec}
\mathbf{y}_{\mathcal{Q}}(t)={\mathcal Q}\left( {\mathbf F}^H {\mathbf X}\left( {\boldsymbol{\theta }} \right) {\mathbf s} \left( t \right) + {\mathbf n}\right).
\end{equation}

\section{DoA Estimation And Channel Estimation}

During pilot sequences transmission, each user transmits $N_p$ pilot symbols simultaneously. Combining $N_p$ quantized vector of \eqref{eq:qyvec} into a matrix yields
\begin{equation}\label{eq:qymtx}
{{\mathbf{Y}}_{\mathcal Q}} = \left[ {{{\mathbf{y}}_{\mathcal Q}}(1), \cdots ,{{\mathbf{y}}_{\mathcal Q}}({N_p})} \right] = {\mathcal Q}\left( {\mathbf F}^H {\mathbf X}\left( {\boldsymbol{\theta }} \right) {\mathbf S} +{\mathbf N} \right), 
\end{equation}
where ${\mathbf S}$ is the $\sum\nolimits_{k = 1}^{{N_t}} {L_k} \times N_p$ pilot matrix with ${\mathbf S} = \left[ {\mathbf s} \left( 1 \right),\cdots,{\mathbf s} \left( N_p \right) \right]$, and ${\mathbf N}$ is the matrix form of ${\mathbf n}$.

By vectorizing ${{\mathbf{Y}}_{\mathcal Q}}$, we obtain
\begin{IEEEeqnarray}{cc}\label{eq:CHqymtx}
{\bf{y}} &= {\rm{vec}}\left( {{{\bf{Y}}_{\cal Q}}} \right) = {\rm{vec}}\left( {{\cal Q}\left( {{{\bf{F}}^H}{\bf{X}}\left( {\boldsymbol{\theta }} \right){\bf{S}} + {\bf{N}}} \right)} \right) \notag \\
& = {\cal Q}\left( {\left( {{{\bf{S}}^T} \otimes {{\bf{F}}^H}} \right){\rm{vec}}\left( {{\bf{X}}\left( {\boldsymbol{\theta }} \right)} \right) + {\rm{vec}}\left( {\bf{N}} \right)} \right),
\end{IEEEeqnarray}
where $\otimes$ denotes the Kronecker product. Defining ${\mathbf A}\triangleq \left( {{{\bf{S}}^T} \otimes {{\bf{F}}^H}} \right)$, ${\underline{\mathbf  x}}\left( {\boldsymbol{\theta }} \right) \triangleq {\rm{vec}}\left( {{\bf{X}}\left( {\boldsymbol{\theta }} \right)} \right)$, and ${\underline{\mathbf n}}\triangleq {\rm{vec}}\left( {\bf{N}} \right)$, we rewrite $\bf y$ as 
\begin{equation}\label{eq:CHqyvec}
{\bf{y}} = {\cal Q}\left( {\mathbf A} {\underline{\mathbf  x}}\left( {\boldsymbol{\theta }} \right) + {\underline{\mathbf n}} \right).
\end{equation}
Therefore, the channel estimation problem is converted to obtaining the sparse vector ${\underline{\mathbf  x}}\left( {\boldsymbol{\theta }} \right)$ from the noisy 1-bit quantized received signal ${\bf{y}}\in {\mathbb C}^{MN_p\times 1}$ under the known linear transform ${\bf A}$.

Since the dimension of ${\bf{X}}\left( {\boldsymbol{\theta }} \right)$ is $M \times \sum_{k=1}^{N_{t}} L_{k}$, $N_t$ and $L_k$ need to be obtained prior to the estimation, which is impractical. To circumvent this issue, we will estimate an $M\times N$ redundant matrix ${\bf{X}}$ instead of ${\bf{X}}\left( {\boldsymbol{\theta }} \right)$ with $N > \sum_{k=1}^{N_{t}} L_{k}$. The estimator $\widehat{\bf X}$ for the redundant matrix ${\bf{X}}$ would be composed of $\widehat{\mathbf  X}\left( {\boldsymbol{\theta }} \right)$  and some vectors whose elements are very close to zero. Hence, by setting a threshold on the norm of each column of $\widehat{\bf X}$, we could get rid of those redundant columns and obtain $\widehat{\mathbf  X}\left( {\boldsymbol{\theta }} \right)$ from $\widehat{\bf X}$. Therefore, \eqref{eq:CHqyvec} can be modified into
\begin{equation}\label{eq:CHqyvecCS}
{\bf{y}} = {\cal Q}\left( {\mathbf A} {{\mathbf  x}} + {\underline{\mathbf n}} \right) = {\cal Q}\left( {\bf z} + {\underline{\mathbf n}} \right),
\end{equation}
where ${{\mathbf  x}} \triangleq {\rm{vec}}\left( {\bf X}\right)$, and we define ${\bf z} \triangleq {\mathbf A} {{\mathbf  x}}$. 

Hence, the channel estimation problem is to recover the sparse vector ${{\mathbf  x}}\in {\mathbb C}^{MN\times 1}$ for given ${\bf{y}} \in {\mathbb C}^{MN_p\times 1}$ and ${\bf A} \in {\mathbb C}^{MN_p\times MN}$. For ease of use, we define $M_y \triangleq MN_p$ and $N_x \triangleq MN$. The CS literature offers a number of high-performance algorithms to recover  ${{\mathbf  x}}$ from ${\bf{y}}$. We would adopt the EM-GAMP algorithm \cite{EMGAMP} to recover ${{\mathbf  x}}$, which extends the GAMP algorithm \cite{GAMP}, for its outstanding performance and computational efficiency.

\subsection{EM-GAMP Algorithm}\label{sec:GAMP}

The EM-GAMP-based algorithm to recover  ${{\mathbf  x}}$ from ${\bf{y}}$ is presented in Algorithm \ref{alg:EMGAMP}, which follows the same structure as EM-GAMP in \cite{EMGAMP} except for the nonlinear steps, e.g., lines \ref{alg:nuz}, \ref{alg:zhat}, \ref{alg:nux} and \ref{alg:xhat} of Algorithm \ref{alg:EMGAMP}. Then, we would focus on describing these nonlinear steps in details for our 1-bit quantization scenario.

First, we compute the posterior variance and mean of $x_m$ in lines \ref{alg:nux}--\ref{alg:xhat} of Algorithm \ref{alg:EMGAMP}. We will suppress the iteration number $t$ for brevity. Following the assumptions in \cite{EMGAMP},  the coefficients $x_n$ in $\bf x$ are assumed to be i.i.d. with Bernoulli Gaussian-mixture (GM) distribution, which yields
\begin{equation}
p_{X}(x_n ; {\bf q})=(1-\lambda) \delta(x_n)+\lambda \sum_{\ell=1}^{L} \omega_{\ell} \mathcal{CN}\left(x_n ; \mu_{\ell}, \psi_{\ell}\right),
\end{equation}
where $\delta(\cdot)$ denotes the Dirac delta function, $L$ is the number of GM components, $\lambda$ is sparse rate, $\omega_{\ell}$, $\mu_{\ell}$ and $\psi_{\ell}$ are the weight, mean, and variance for the $\ell$th GM component respectively. The prior parameters ${\bf q} \triangleq [\lambda,\omega_{1},\cdots,\omega_{L},\mu_{1},\cdots,\mu_{L},\psi_{1},\cdots,\psi_{L}]$ are unknown but will be learned and updated by EM algorithm at each iteration. Moreover, $L$ are treated as fixed and known. Under these assumptions, the conditional PDF $p_{X|Y}$ in \eqref{eq:cdpdfx}, the posterior variance and mean of $x_m$ in lines \ref{alg:nux}--\ref{alg:xhat}  are given\footnote{The multiplication rule for Gaussian PDF:  $\mathcal{CN}(x ; a, A) \mathcal{CN}(x ; b, B) = \mathcal{CN}\left(x ; \frac{\alpha / A+b / B}{1 / A+1 / B}, \frac{1}{1 / A+1 / B}\right)\mathcal{CN}\left(0 ; a-b, A+B\right)$ is utilized in the derivations.}
in \cite{EMGAMP}.

\begin{algorithm}[!tbp]
\caption{The EM-GAMP-based Algorithm}
\label{alg:EMGAMP}
\begin{algorithmic}[1]
\Require $p_{X}(\cdot)$, $p_{Y | Z}(\cdot | \cdot)$, ${\bf y}\in {\mathbb C}^{M_y\times 1}$, ${\bf A}\in {\mathbb C}^{M_y\times N_x}$, $T_{\max }$
\State {\bf define:}
\begin{equation}\label{eq:cdpdfz}
p_{Z|Y}\left(z_{m} | y_{m}; \widehat{p}_{m} , \nu^{p}_{m}\right) \triangleq \frac{p_{Y|Z}\left(y_{m} | z_{m}\right) \mathcal{C} \mathcal{N}\left(z_{m} ; \widehat{p}_{m}, \nu^{p}_{m}\right)}{\int_{z} p_{Y | Z}\left(y_{m} | z\right) \mathcal{C} \mathcal{N}\left(z ; \widehat{p}_{m}, \nu^{p}_{m}\right)}
\end{equation}
\begin{equation}\label{eq:cdpdfx}
p_{X | \mathbf{Y}}\left(x_{n} | {\bf y}; \widehat{r}_{n} , \nu^{r}_{n}, {\bf q}\right) \triangleq \frac{p_{X}\left(x_{n} ; {\bf q}\right)  \mathcal{CN}\left(x_{n} ; \widehat{r}_{n}, \nu^{r}_{n}\right)}{\int_{x} p_{X}(x ; {\bf q}) \mathcal{CN}\left(x ; \widehat{r}_{n}, \nu^{r}_{n}\right)}
\end{equation}
\State {\bf initialize:}
\Statex {\begin{center} $ {\bf q} $ \end{center} } 
\Statex {\begin{center} $ \forall m \in [1,\cdots,M_y]: \widehat{s}_{m}(0)=0 $ \end{center} } 
\Statex {\begin{center} $ \forall n \in [1,\cdots,N_x]: \widehat{x}_{n}(1)=\int_{x} x \cdot p_{X}(x;{\bf q}) $ \end{center} } 
\Statex {\begin{center} $ \forall n : \nu_{n}^{x}(1)=\int_{x}\left|x-\widehat{x}_{n}(1)\right|^{2} p_{X}(x;{\bf q}) $ \end{center} }
\For{$t=1:T_{\max }$}
\State  $\forall m \in [1,M_y]: \nu_{m}^{p}(t)=\sum_{n=1}^{N_x}\left|A_{m n}\right|^{2} \nu_{n}^{x}(t)$
\State $\forall m : \widehat{p}_{m}(t)=\sum_{n=1}^{N_x} A_{m n} \widehat{x}_{n}(t)-\nu_{m}^{p}(t) \widehat{s}_{m}(t-1)$
\State \label{alg:nuz} $\forall m : \nu_{m}^{z}(t)=\operatorname{var}_{Z | Y}\left\{z_{m} | y_{m} ; \widehat{p}_{m}(t), \nu_{m}^{p}(t)\right\}$
\State \label{alg:zhat} $\forall m : \widehat{z}_{m}(t)=\mathbb{E}_{Z | Y}\left\{z_{m} | y_{m} ; \widehat{p}_{m}(t), \nu_{m}^{p}(t)\right\}$
\State $\forall m: \nu_{m}^{s}(t)=\left(1-\nu_{m}^{z}(t) / \nu_{m}^{p}(t)\right) / \nu_{m}^{p}(t)$
\State $\forall m : \widehat{s}_{m}(t)=\left(\widehat{z}_{m}(t)-\widehat{p}_{m}(t)\right) / \nu_{m}^{p}(t)$
\State $\forall n \in [1,N_x]: \nu_{n}^{r}(t)=\left(\sum_{m=1}^{M_y}\left|A_{m n}\right|^{2} \nu_{m}^{s}(t)\right)^{-1}$
\State $\forall n : \widehat{r}_{n}(t)=\widehat{x}_{n}(t)+\nu_{n}^{r}(t) \sum_{m=1}^{M_y} A_{m n}^{*} \widehat{s}_{m}(t)$
\State \label{alg:nux}$\forall n : \nu_{n}^{x}(t+1)=\operatorname{var}_{X | Y}\left\{x_{n} | {\bf y} ; \widehat{r}_{n}(t), \nu_{n}^{r}(t), {\bf q}\right\}$
\State \label{alg:xhat}$\forall n : \widehat{x}_{n}(t+1)=\mathbb{E}_{X | Y}\left\{x_{n} | {\bf y} ; \widehat{r}_{n}(t), \nu_{n}^{r}(t), {\bf q}\right\}$
\State update $\bf q$ by using EM algorithm,
\EndFor
\Ensure $\widehat{\bf x}$
\end{algorithmic} 
\end{algorithm}

Next, we derive the expressions of the conditional PDF $p_{Z|Y}$, the posterior variance and mean of $z_m$ in lines \ref{alg:nuz}--\ref{alg:zhat} of Algorithm \ref{alg:EMGAMP}. Since the 1-bit quantizers are applied separately to the real and imaginary parts of the complex inputs, the derivation should be done for real and imaginary parts respectively. Due to the noise in \eqref{eq:CHqyvecCS} is circular symmetric Gaussian, we only need to derive the expressions of the real parts, while the expressions for the imaginary parts can be obtained analogously. For ease of notation, we will abuse $y$, $z$ and $\widehat{p}$ to denote $\Re\{y\}$, $\Re\{z\}$ and $\Re\{\widehat{p}\}$ and suppress the iteration number $t$ in the following derivations.

Note that in \eqref{eq:CHqyvecCS}, $\underline{\bf n}$ follows $\mathcal{CN}\left(\underline{\mathbf{n}}; 0, \sigma^2 \mathbf{I}\right)$  and $\mathcal{Q}(\cdot)$ denotes 1-bit quantizer, which yields
\begin{equation}\label{eq:1bcdpdf}
p_{Y|Z}\left(y_{m} | z_{m}\right) = \Phi\left(\frac{\operatorname{sign}\left(y_{m}\right)z_{m}}{\sigma/\sqrt{2}}\right),
\end{equation}
where $\Phi\left(\cdot\right)$ denotes the CDF of the standard normal distribution. Plugging \eqref{eq:1bcdpdf} into \eqref{eq:cdpdfz} and plugging \eqref{eq:cdpdfz} into line \ref{alg:zhat} of Algorithm \ref{alg:EMGAMP} yields\footnote{The facts that $\int_{-\infty}^{\infty} \Phi(a+b x) \phi(x) d x=\Phi\left(\frac{a}{\sqrt{1+b^{2}}}\right)$ and $\int_{-\infty}^{\infty} x \Phi(a+b x) \phi(x) d x=\frac{b}{\sqrt{1+b^2}} \phi\left(\frac{a}{\sqrt{1+b^2}}\right)$ are applied in the derivations.}
\begin{IEEEeqnarray}{ll}
\widehat{z}_{m} &= \frac{\int_{z} z \cdot p_{Y|Z}\left(y_{m} | z\right) \cdot \mathcal{N}\left(z ; \widehat{p}_{m}, \nu^{p}_{m}/2\right)}{\int_{z} p_{Y | Z}\left(y_{m} | z\right) \cdot \mathcal{N}\left(z ; \widehat{p}_{m}, \nu^{p}_{m}/2\right)} \notag\\
&=\widehat{p}_{m}+\frac{\operatorname{sign}\left(y_{m}\right) \frac{\nu^{p}_{m}}{2}}{\sqrt{{\left(\sigma^{2}+\nu^{p}_{m}\right)}/{2}}} \frac{\phi\left(\eta_{m}\right)} {\Phi\left(\eta_{m}\right)},
\end{IEEEeqnarray}
where $\phi\left(\cdot\right)$ denotes the PDF of the standard normal distribution and
\begin{equation}
\eta_{m}= \frac{\operatorname{sign}(y_{m}) \widehat{p}_{m}}{\sqrt{{\left(\sigma^{2}+\nu^{p}_{m}\right)}/{2}} }.
\end{equation}
Similarly, substituting \eqref{eq:1bcdpdf} into \eqref{eq:cdpdfz} and plugging \eqref{eq:cdpdfz} into line \ref{alg:nuz} of Algorithm \ref{alg:EMGAMP} yields\footnote{The fact that $\int_{-\infty}^{\infty} x^2 \Phi(a+b x) \phi(x) d x = \Phi\left(\frac{a}{t}\right) - \frac{a b^2}{t^3} \phi\left(\frac{a}{t}\right)$ is applied in the derivations, where $t=\sqrt{1+b^2}$.}
\begin{IEEEeqnarray}{ll}
\nu^{z}_{m} &= \frac{\int_{z} (z-\widehat{z}_{m})^2 \cdot p_{Y|Z}\left(y_{m} | z\right) \cdot \mathcal{N}\left(z ; \widehat{p}_{m}, \nu^{p}_{m}/2\right)}{\int_{z} p_{Y | Z}\left(y_{m} | z\right) \cdot \mathcal{N}\left(z ; \widehat{p}_{m}, \nu^{p}_{m}/2\right)} \notag\\
&=\frac{\nu^p_{m}}{2} - \frac{ \left(\frac{\nu^{p}_{m}}{2}\right)^2 }{\frac{\sigma^{2}}{2} + \frac{\nu^{p}_{m}}{2}} \frac{\phi\left(\eta_{m}\right)} {\Phi\left(\eta_{m}\right)}  \left(\eta_{m} + \frac{\phi\left(\eta_{m}\right)} {\Phi\left(\eta_{m}\right)}\right).
\end{IEEEeqnarray}
Note that the above derivations could be extended to the case of few-bits quantization. Further details about few-bits quantization can be found in \cite{JDC}.

\subsection{Gridless DoA Estimation}\label{sec:DoA}

Applying Algorithm \ref{alg:EMGAMP} and setting thresholds, we then obtain $\widehat{\bf X}\left( {\boldsymbol{\theta }} \right)$ which denotes the estimator of ${\bf X}\left( {\boldsymbol{\theta }} \right)$. According to \eqref{eq:effCH}, the IDFT of  $\widehat{\bf X}\left( {\boldsymbol{\theta }} \right)$ is the estimator of the Vandermonde matrix ${\bf V}\left( {\boldsymbol{\theta }} \right){\mathbf D}$ of which each column corresponds to one DoA. We need to estimate all the corresponding DoA from ${\bf F}^{H}\widehat{\bf X}\left( {\boldsymbol{\theta }} \right)$. We use $\widehat{\boldsymbol{v}}_{k}$ to present the $k$th column of ${\bf F}^{H}\widehat{\bf X}\left( {\boldsymbol{\theta }} \right)$, and the $k$th column of ${\bf V}\left( {\boldsymbol{\theta }} \right){\mathbf D}$ is $\alpha_{k} \mathbf{a}_{r}\left(\theta_{ k}\right)$. Assume that $\widehat{\boldsymbol{v}}_{k}$ is a measurement of $\alpha_{k} \mathbf{a}_{r}\left(\theta_{ k}\right)$ under the inference of the AWGN noise $\mathbf{w}$ with $\mathbf{w}\sim\mathcal{CN}(\mathbf{w};\mathbf{0},\sigma_{w}^2\mathbf{I})$. We obtain
\begin{equation}
\widehat{\boldsymbol{v}}_{k} = \alpha_{k} \mathbf{a}_{r}\left(\theta_{ k}\right) + \mathbf{w}.
\end{equation}
Given the measurement $\widehat{\boldsymbol{v}}_{k}$, the likelihood function of $\theta_{ k}$ and  $\alpha_{k}$ is 
\begin{IEEEeqnarray}{ll}
\ln p\left( \widehat{\boldsymbol{v}}_{k}; \theta_{ k},\alpha_{k}\right) &=\ln\left( \frac{1}{\pi^M\sigma_{w}^{2M}} e^{-\frac{\Vert \alpha_{k} \mathbf{a}_{r}\left(\theta_{k}\right) - \widehat{\boldsymbol{v}}_{k} \Vert^{2}_{2}}{\sigma_{w}^{2}}}\right).
\end{IEEEeqnarray}
The ML estimators for $\alpha_{k}$ and $\theta_{k}$ can then be obtained from
\begin{IEEEeqnarray}{lcl}
\left(\widehat{\alpha}_{k},\widehat{\theta}_{k}\right)_{\mathrm{ML}}&=\argmax\limits_{\alpha_{k},\theta_{k}} &\ln p\left( \widehat{\boldsymbol{v}}_{k}; \theta_{ k},\alpha_{k}\right). \label{eq:MLDoA}
\end{IEEEeqnarray}
We find, after some algebra, that \eqref{eq:MLDoA} is equivalent to
\begin{IEEEeqnarray}{l}\label{eq:DoA}
\widehat{\theta}_{k,\mathrm{ML}} = \argmin\limits_{\theta_{k}} \Vert \mathbf{a}^{\dag}_{r}\left(\theta_{k}\right) \widehat{\boldsymbol{v}}_{k} \mathbf{a}_{r}\left(\theta_{k}\right) - \widehat{\boldsymbol{v}}_{k} \Vert^{2}_{2}, \\
\widehat{\alpha}_{k,\mathrm{ML}} = \mathbf{a}^{\dag}_{r}\left(\widehat{\theta}_{k,\mathrm{ML}}\right)\widehat{\boldsymbol{v}}_{k},
\end{IEEEeqnarray}
where  $\dag$ denotes the pseudo inverse of a matrix. Therefore, all the estimates $\widehat{\boldsymbol{\theta}}$ of the DoAs can be obtained by solving \eqref{eq:DoA} for all $k$.

Note that the ML estimation for $\widehat{\theta}_{k}$ in \eqref{eq:MLDoA} needs to be achieved by 1-D exhaustive search, which is undesirable in some applications. Thus, we here provide an alternative method. Since there is only one snapshot $\widehat{\boldsymbol{v}}_{k}$ and only one DoA $\widehat{\theta}_{k}$ needs to estimate, we could apply the search-free root-MUSIC algorithm or ESPRIT algorithm or the computationally efficient Newton's method in \cite{cqian} to $\widehat{\boldsymbol{v}}_{k}$. 

\subsection{Channel Estimation}\label{sec:CH} 

In the above section, we obtain the $\widehat{\bf X}\left( {\boldsymbol{\theta }} \right)$ and  $\widehat{\boldsymbol{\theta}}$ which denote the estimates of ${\bf X}\left( {\boldsymbol{\theta }} \right)$ and ${\boldsymbol{\theta}}$ respectively. Recalling from \eqref{eq:effCH} that ${\bf{H}}= {\bf{F}}^H {\mathbf X}\left( {\boldsymbol{\theta }} \right)$ and $  {\bf X}\left( {\boldsymbol{\theta }} \right) ={\bf{F}}{\bf{V}}\left( {\boldsymbol{\theta }} \right){\bf{D}} $, we see that the estimator $\widehat{\bf{H}}$ of the effective channel ${\bf{H}}$ is 
\begin{equation}\label{eq:estiH}
\widehat{\bf{H}} = {\bf{F}}^H \widehat{\mathbf X}\left( {\boldsymbol{\theta }} \right),
\end{equation}
where $\widehat{\bf X}\left( {\boldsymbol{\theta }} \right)$ is from Section \ref{sec:GAMP} Algorithm \ref{alg:EMGAMP}. The estimate $\widehat{\bf{D}}$ of the channel complex gain matrix ${\bf{D}}$ can be obtained by using LS method, which yields
\begin{equation}\label{eq:estiD}
\widehat{\bf{D}} = {\bf{V}}\left( \widehat{\boldsymbol{\theta }} \right)^{\dag } {\bf{F}}^H \widehat{\mathbf X}\left( {\boldsymbol{\theta }} \right).
\end{equation}

\subsection{Discussions}\label{sec:discuss}

Since the 1-bit quantizers keep only the sign of the real and imaginary parts of the inputs, the information of the amplitude of the inputs is lost after quantization especially in the high SNR regime. Therefore, the effective channel $\mathbf{H}$ and channel complex gain matrix $\mathbf{D}$ can only be reconstructed up to a real scaling factor, as also discussed in \cite{BIHTDoA,MJH}. In order to reconstruct $\mathbf{H}$ and $\mathbf{D}$, the norm estimation of the received unquantized signal should be considered.  Recalling from \eqref{eq:CHqyvecCS} that the unquantized signal is $\mathbf {A x}+\underline{\mathbf{n}}$, we follow the  assumption in \cite{MJH} that the total received power of $\mathbf {A x}+\underline{\mathbf{n}}$ can be measured before quantization. Then we get 
\begin{equation}
\mathbb{E}\left\{\|\mathbf {A x}+\underline{\mathbf{n}}\|^{2}\right\} = \left\|\mathbf{A}\right\|_{F}^{2} \sigma_{x}^{2}+N_{p} M \sigma^{2},
\end{equation} 
where $\|\cdot\|_{F}$ denotes the Frobenius norm. The variance $\sigma_{x}^{2}$ then can written as 
\begin{equation}
\sigma_{x}^{2} = \frac{\mathbb{E}\left\{\|\mathbf {A x}+\underline{\mathbf{n}}\|^{2}\right\} - N_{p} M \sigma^{2}}{\left\|\mathbf{A}\right\|_{F}^{2} }. 
\end{equation} 
From the law of large numbers, we know that $\|\mathbf{x}\| \approx \sigma_{x} \sqrt{M N} $. After obtaining $\widehat{\mathbf{x}}$ from Algorithm \ref{alg:EMGAMP}, we scale $\widehat{\mathbf{x}}$ as
\begin{equation}\label{eq:normalize}
\widetilde{\mathbf{x}} = \frac{\widehat{\mathbf{x}}}{\| \widehat{\mathbf{x}} \|}\|\mathbf{x}\|.
\end{equation}

\begin{algorithm}[!t]
\caption{The Framework of The Proposed Algorithm}
\label{alg:proposed}
\begin{algorithmic}[1]
\Require  $\mathbf{y}$, ${\bf A}$
\State Apply EM-GAMP algorithm in Algorithm \ref{alg:EMGAMP} to obtain $\widehat{\bf x}$.
\State Estimate the norm of  the unquantized signal, normalize $\widehat{\bf x}$ according to \eqref{eq:normalize} and set thresholds to obtain $\widehat{\bf X}(\boldsymbol{\theta})$.
\State Estimate all the decoupled DoAs separately from $\widehat{\bf X}(\boldsymbol{\theta})$ by using ML estimator or root-MUSIC or ESPRIT.
\State Estimate the channel coefficients and gains from $\widehat{\bf X}(\boldsymbol{\theta})$ according to  \eqref{eq:estiH} and \eqref{eq:estiD}.
\Ensure $\widehat{\boldsymbol{\theta}}$, $\widehat{\mathbf{H}}$ and $\widehat{\mathbf{D}}$
\end{algorithmic} 
\end{algorithm}

Hence, summing  all the above discussions up, the framework of the proposed algorithm is presented in Algorithm \ref{alg:proposed}.

\section{Simulation Results}
The performances  of the proposed EM-GAMP-based algorithm on 1-bit DoA estimation and channel estimation are evaluated in this section. The pilot length is chosen as $N_p=16$. The pilot sequence is chosen as a length-$N_p$ Zadoff-Chu (ZC) sequence, and each row of the pilot matrix $\mathbf{S}$ is a circularly shifted version of the ZC sequence. This assumption would assure that the pilot sequences corresponding to different DoAs are orthogonal and would address the rank deficiency issue when DoAs are extremely close. The angular domain channel parameters are generated randomly once and fixed at all times with 
\begin{itemize}
\item DoAs vector $\boldsymbol{\theta}=[-40^{\circ},20^{\circ},20.005^{\circ}]$  where two DoAs are extremely close,
\item channel complex gains matrix ${\bf{D}} = {\rm{diag}}\{ 0.8084 + 0.5887\jmath, 0.6884 + 0.7254\jmath, 0.7344 - 0.6787\jmath \}$.
\end{itemize}

The SNR  is defined as 
\begin{equation}
\mathrm{SNR} \triangleq 10\log_{10}\left( \frac{\| \mathbf{S} \|^{2}_{F}}{N_p \sigma^2} \right)\quad \mathrm{dB}.
\end{equation}
The noise variance is assumed to be $\sigma^2 = 1$. The performance metric of DoA estimation is the root mean square error (RMSE), and the performance of channel gains estimator is measured in terms of the normalized mean square error (NMSE), which are defined as
\begin{IEEEeqnarray}{l}
\mathrm{RMSE}_{\boldsymbol{\theta}} \triangleq \sqrt{ \mathbb{E}\left\{ \frac{\| \widehat{ \boldsymbol{\theta}}  - \boldsymbol{\theta} \|^2_2}{\sum_{k=1}^{N_{t}} L_{k}} \right\} }, \, \mathrm{NMSE}_{\mathbf{D}} \triangleq \mathbb{E}\left\{\frac{\|\widehat{\mathbf{D}}-\mathbf{D}\|^{2}_{F}}{\|\mathbf{D}\|^{2}_{F}}\right\}. \notag
\end{IEEEeqnarray}

In the following simulation, we would compare our EM-GAMP-based method\footnote{The proposed EM-GAMP-based algorithm is implemented based on the open-source "GAMPmatlab" software suite which is available at \href{http://sourceforge.net/projects/gampmatlab/} {http://sourceforge.net/projects/gampmatlab/.}} with the BIHT algorithm proposed in \cite{BIHT}, the LASSO problem and the BPDN problem, where the LASSO and BPDN problems are solved by using the software in \cite{SPGL1}.

\begin{figure}[!tbp]
\centering
\includegraphics[width=0.38\textwidth]{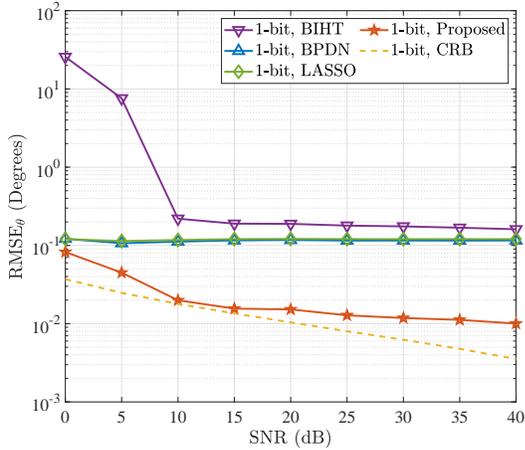}
\caption{$\mathrm{RMSE}_{\boldsymbol{\theta}}$ versus SNR for $M=90$ and $\boldsymbol{\theta}=[-40^{\circ},20^{\circ},20.005^{\circ}]$, where two DoAs are extremely close.}
\label{fig:narrowRMSESNR}
\end{figure}

\begin{figure}[!tbp]
\centering
\includegraphics[width=0.38\textwidth]{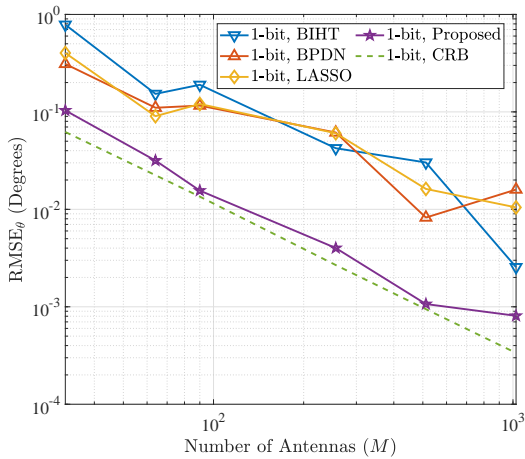}
\caption{$\mathrm{RMSE}_{\boldsymbol{\theta}}$ versus $M$ for $\mathrm{SNR}=15\,\mathrm{dB}$ and extremely close DoAs with $\boldsymbol{\theta}=[-40^{\circ},20^{\circ},20.005^{\circ}]$.}.
\label{fig:narrowRMSEM}
\end{figure}

\begin{figure}[!tbp]
\centering
\includegraphics[width=0.38\textwidth]{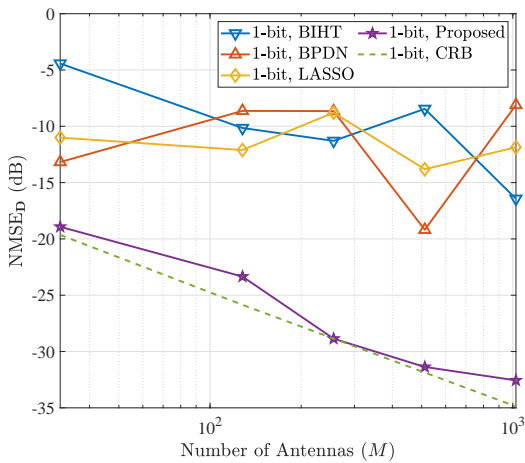}
\caption{$\mathrm{NMSE}_{\mathbf{D}}$ versus $M$ for $\mathrm{SNR}=15\,\mathrm{dB}$ and very close DoAs with $\boldsymbol{\theta}=[-40^{\circ},20^{\circ},20.005^{\circ}]$.}
\label{fig:narrowNMSEM}
\end{figure}

Fig. \ref{fig:narrowRMSESNR} shows the RMSE of DoA estimation  versus SNR for $M=90$ antennas. For the case of 1-bit quantization, the  proposed method outperforms all other algorithms in distinguishing very close DoAs because all the DoAs are decoupled and estimated separately. Additionally,  a nonzero estimation error floor arises due to 1-bit quantization and cannot be eliminated by increasing SNR, as also discussed in \cite{XLY}. The CRBs of $\boldsymbol{\theta}$ are provided as benchmarks, which are derived in \cite{1bRELAX}. Fig. \ref{fig:narrowRMSEM} shows the RMSE of DoA estimation  versus $M$ for for $\mathrm{SNR}=15\,\mathrm{dB}$. The  proposed method surpasses all other algorithms for 1-bit quantization. In Fig. \ref{fig:narrowNMSEM}, the NMSE of channel gains estimation is plotted  over $M$ for $\mathrm{SNR}=15\,\mathrm{dB}$. The  proposed method surpasses all other algorithms for the case of 1-bit quantization, which shows the advance and robustness of the proposed method. 

\section{Conclusion}

In this paper, based on EM-GAMP algorithm, we proposed a gridless DoA estimation and channel estimation methodology for 1-bit quantized mmWave massive MIMO system. The prior knowledge about the number of DoAs is unnecessary for the proposed method. In addition, the proposed algorithm outperforms all other algorithms in distinguishing very close DoAs since all the DoAs are decoupled and estimated separately. The discussions support the feasibility of applying the proposed algorithm in  practical one-bit quantized mmWave massive MIMO systems.

\balance
\bibliographystyle{IEEEtran}
\input{GCpaper.bbl}

\end{document}

%% file: GCpaper.bbl

%% file: GCpaper.bbl
\begin{thebibliography}{10}
\providecommand{\url}[1]{#1}
\csname url@samestyle\endcsname
\providecommand{\newblock}{\relax}
\providecommand{\bibinfo}[2]{#2}
\providecommand{\BIBentrySTDinterwordspacing}{\spaceskip=0pt\relax}
\providecommand{\BIBentryALTinterwordstretchfactor}{4}
\providecommand{\BIBentryALTinterwordspacing}{\spaceskip=\fontdimen2\font plus
\BIBentryALTinterwordstretchfactor\fontdimen3\font minus
  \fontdimen4\font\relax}
\providecommand{\BIBforeignlanguage}[2]{{%
\expandafter\ifx\csname l@#1\endcsname\relax
\typeout{** WARNING: IEEEtran.bst: No hyphenation pattern has been}%
\typeout{** loaded for the language `#1'. Using the pattern for}%
\typeout{** the default language instead.}%
\else
\language=\csname l@#1\endcsname
\fi
#2}}
\providecommand{\BIBdecl}{\relax}
\BIBdecl

\bibitem{MIMODoA}
F.~{Boccardi}, R.~W. {Heath}, A.~{Lozano}, T.~L. {Marzetta}, and P.~{Popovski},
  ``Five disruptive technology directions for {5G},'' \emph{IEEE Commun. Mag.},
  vol.~52, no.~2, pp. 74--80, 2014.

\bibitem{BIHTDoA}
C.~{St{\"o}ckle}, J.~{Munir}, A.~{Mezghani}, and J.~A. {Nossek}, ``1-bit
  direction of arrival estimation based on compressed sensing,'' in \emph{Proc.
  IEEE 16th Int. Workshop Signal Processing Adv. in Wireless Commun. (SPAWC)},
  2015, pp. 246--250.

\bibitem{SVMDoA}
Y.~{Gao}, D.~{Hu}, Y.~{Chen}, and Y.~{Ma}, ``Gridless 1-b {DOA} estimation
  exploiting {SVM} approach,'' \emph{IEEE Commun. Lett.}, vol.~21, no.~10, pp.
  2210--2213, 2017.

\bibitem{WenDoA}
C.~{Wang}, C.~{Wen}, S.~{Jin}, and S.~{Tsai}, ``Gridless channel estimation for
  mixed one-bit antenna array systems,'' \emph{IEEE Trans. Wireless Commun.},
  vol.~17, no.~12, pp. 8485--8501, Dec. 2018.

\bibitem{BIHT}
L.~{Jacques}, J.~N. {Laska}, P.~T. {Boufounos}, and R.~G. {Baraniuk}, ``Robust
  1-bit compressive sensing via binary stable embeddings of sparse vectors,''
  \emph{IEEE Trans. Inf. Theory}, vol.~59, no.~4, pp. 2082--2102, 2013.

\bibitem{1bRELAX}
F.~{Liu}, H.~{Zhu}, J.~{Li}, P.~{Wang}, and P.~V. {Orlik}, ``Massive {MIMO}
  channel estimation using signed measurements with antenna-varying
  thresholds,'' in \emph{Proc. IEEE Statistical Signal Process. Workshop},
  2018, pp. 188--192.

\bibitem{WBL}
B.~{Wang}, F.~{Gao}, S.~{Jin}, H.~{Lin}, and G.~Y. {Li}, ``Spatial- and
  frequency-wideband effects in millimeter-wave massive {MIMO} systems,''
  \emph{IEEE Trans. Signal Process.}, vol.~66, no.~13, pp. 3393--3406, 2018.

\bibitem{xhxTVT}
H.~{Xie}, F.~{Gao}, S.~{Zhang}, and S.~{Jin}, ``A unified transmission strategy
  for {TDD/FDD} massive {MIMO} systems with spatial basis expansion model,''
  \emph{IEEE Trans. Veh. Technol.}, vol.~66, no.~4, pp. 3170--3184, 2017.

\bibitem{EMGAMP}
J.~P. {Vila} and P.~{Schniter}, ``Expectation-maximization {Gaussian}-mixture
  approximate message passing,'' \emph{IEEE Trans. Signal Process.}, vol.~61,
  no.~19, pp. 4658--4672, 2013.

\bibitem{GAMP}
S.~{Rangan}, ``Generalized approximate message passing for estimation with
  random linear mixing,'' in \emph{Proc. IEEE Int. Symp. Inf. Theory (ISIT)},
  2011, pp. 2168--2172.

\bibitem{JDC}
C.~{Wen}, C.~{Wang}, S.~{Jin}, K.~{Wong}, and P.~{Ting}, ``Bayes-optimal joint
  channel-and-data estimation for massive {MIMO} with low-precision {ADCs},''
  \emph{IEEE Trans. Signal Process.}, vol.~64, no.~10, pp. 2541--2556, May
  2016.

\bibitem{cqian}
C.~{Qian}, ``A simple modification of {ESPRIT},'' \emph{IEEE Signal Process.
  Lett.}, vol.~25, no.~8, pp. 1256--1260, Aug. 2018.

\bibitem{MJH}
J.~{Mo}, P.~{Schniter}, and R.~W. {Heath}, ``Channel estimation in broadband
  millimeter wave {MIMO} systems with few-bit {ADCs},'' \emph{IEEE Trans.
  Signal Process.}, vol.~66, no.~5, pp. 1141--1154, 2018.

\bibitem{SPGL1}
E.~Van Den~Berg and M.~P. Friedlander, ``Probing the pareto frontier for basis
  pursuit solutions,'' \emph{SIAM J. Sci. Comput.}, vol.~31, no.~2, pp.
  890--912, 2009.

\bibitem{XLY}
L.~Xu, X.~Lu, S.~Jin, F.~Gao, and Y.~Zhu, ``On the uplink achievable rate of
  massive {MIMO} system with low-resolution {ADC} and {RF} impairments,''
  \emph{IEEE Commun. Lett.}, vol.~23, no.~3, pp. 502--505, 2019.

\end{thebibliography}
